# Optical Manipulation of Single Flux Quanta

I. S. Veshchunov[1,2,4 §], W. Magrini[1,2,3 §], S. V. Mironov[3,4], A. G. Godin[1,2], J.-B. Trebbia[1,2], A. I. Buzdin[3], Ph. Tamarat,[1,2] and B. Lounis[1,2] *

[1] Université de Bordeaux, LP2N, F-33405 Talence, France
[2] Institut d'Optique & CNRS, LP2N, F-33405 Talence, France
[3] Université de Bordeaux, LOMA, F-33405 Talence, France
[4] Moscow Institute of Physics and Technology, 141700 Dolgoprudny, Russia

§ Equal contributions
* brahim.lounis@u-bordeaux.fr

**Magnetic field can penetrate into type-II superconductors in the form of Abrikosov vortices, which are magnetic flux tubes surrounded by circulating supercurrents[1] often trapped at defects referred to as pinning sites. Although the average properties of the vortex matter can be tuned with magnetic fields[2], temperature or electric currents[3], handling of individual vortices remains challenging and has been demonstrated only with sophisticated magnetic force[4,5], superconducting quantum interference device[6,7] or strain-induced[8] scanning local probe microscopies. Here, we introduce a far-field optical method based on local heating of the superconductor with a focused laser beam to realize a fast, precise and non-invasive manipulation of individual Abrikosov vortices, in the same way as with optical tweezers. This simple approach provides the perfect basis for sculpting the magnetic flux profile in superconducting devices like a vortex lens or a vortex cleaner, without resorting to static pinning or ratchet effects. Since a single vortex can induce a Josephson phase shift[9], our method also paves the way to fast optical drive of Josephson junctions, with potential massive parallelization of operations.**

Although the possibility to induce a global vortex flow by thermal gradients in superconductors (SC) was experimentally demonstrated almost 50 years ago[10,11], thermal manipulation of individual Abrikosov vortices has not yet been achieved. The early observations of vortex flows in weak thermal gradients were interpreted on the basis that vortices behave as entropy carrying particles, which obey general thermal diffusion laws and therefore are in search of colder places in the superconductor. However, theoretical investigations revisiting the problem of entropy transport in superconductors show that the vortex superconducting current does not carry entropy[12]. It was suggested that the thermal force acting on a vortex is directed towards the hot region of the superconductor. Here we experimentally demonstrate that the immersion of an isolated vortex in a strong thermal gradient generates a force, which attracts the vortex to the hot region of the superconductor, and use it as a basis for optical manipulation of single flux quanta.

The free energy per unit length of an isolated vortex is proportional to the density of Cooper pairs and is given by:

$$U \approx \frac{\Phi_0^2}{4\pi\mu_0\lambda^2}\ln\left(\frac{\lambda}{\xi}\right) \qquad (1),$$

where $\Phi_0$ is the flux quantum, $\mu_0$ the vacuum permeability, $\lambda$ the London magnetic penetration depth and $\xi$ the coherence length. It linearly grows as $T_c - T$ when reducing the temperature $T$ down from the superconducting critical temperature $T_c$[13,14]. As a consequence, a temperature gradient in the superconductor will generate a thermal force per vortex unit length given by

$$\boldsymbol{F} \approx \frac{\Phi_0^2}{4\pi\mu_0\lambda_0^2}\ln\left(\frac{\lambda_0}{\xi_0}\right)\frac{\boldsymbol{\nabla}T}{T_c} \qquad (2),$$

where $\lambda_0$ and $\xi_0$ are the values of $\lambda$ and $\xi$ at zero temperature. In this work, a laser beam tightly focused to a diffraction-limited spot locally heats the superconductor and creates a micron-sized "hotspot" with a temperature rise in the Kelvin range, while keeping the temperature below $T_c$. The large thermal gradient of the order of the K/$\mu$m can easily be tuned with laser power, so that the generated thermal force overcomes the pinning potential and induces a vortex motion towards the laser focus. Therefore the laser beam acts as optical tweezers that move single flux quanta to any new desired positions in the superconductor.

The experimental setup used for single vortex optical manipulation and magneto-optical imaging is depicted in Fig. 1a. A 90 nm-thick niobium film was cooled below its critical temperature $T_c = 8.6$ K under a weak external magnetic field $H_{ext}$ applied perpendicular to the niobium film, in order to set the SC in the mixed state. Fig. 1b is a magneto-optical image of a spontaneous vortex distribution obtained at $T = 4.6$ K and for $H_{ext} = 0.024$ Oe. The vortices are located at pinning sites that are randomly distributed in the SC sample. Vortex after vortex, they are then optically dragged and repositioned into an artificial pattern forming the letters A V for "Abrikosov Vortices", as shown in Fig. 1c.

Because they dissipate energy and generate internal noise, vortices constitute a serious obstacle limiting the operation of numerous superconducting devices[15]. The most desirable method to overcome this difficulty would be to remove the vortices from the bulk of the superconductor without the need for material structuration or the incorporation of impurities and defects[16,17]. To this purpose, we show in Figures 1d-e how a vortex-free area is produced in a niobium film by simply scanning a focused laser beam which picks up the vortices in its path, drags and drops them in a bordering area of the superconductor, like a "vortex-broom". Actually, because of their mutual repulsion, all vortices cannot be piled up in the same optical spot. This explains the "comet-like" shape of the vortex distribution in the region along the laser path, with a maximal vortex density at the final position of the laser.

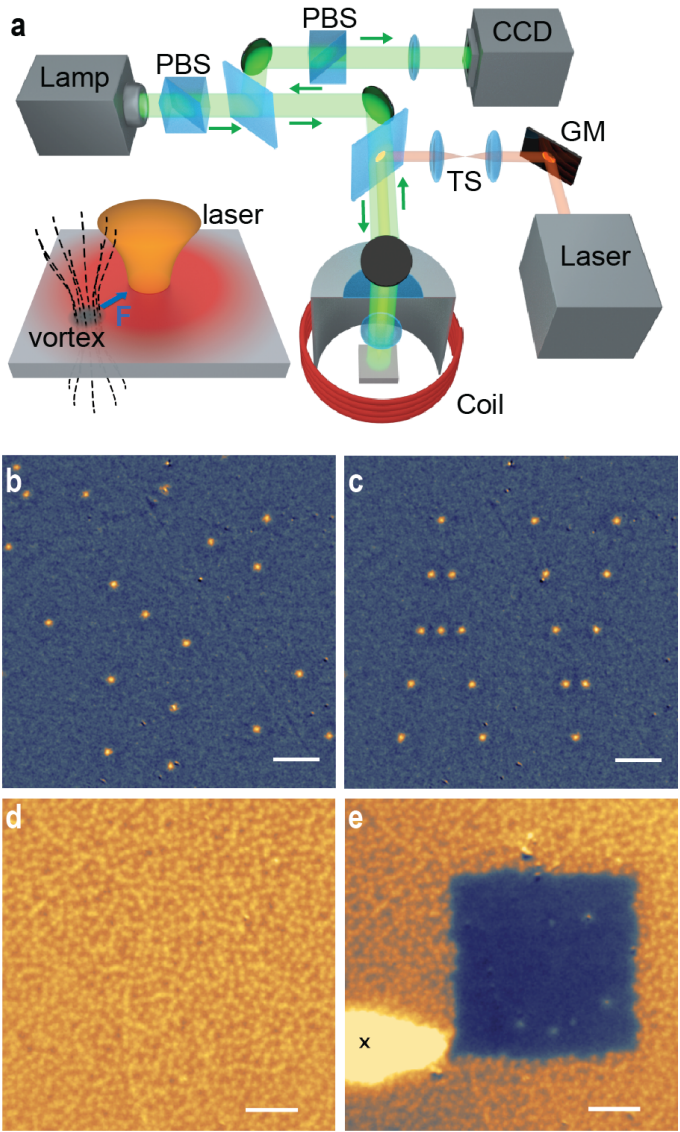

**Figure 1 : Single vortex manipulation with a focused laser beam.**

**a**, The concept of vortex attraction in a thermal gradient induced by a laser spot is illustrated. Magneto-optical imaging of individual vortices is based on the Faraday rotation of light polarization in a Bi:LuIG garnet layer placed onto the superconductor, in a crossed-polarizer beam path configuration[18-20] (PBS: polarizing beam-splitter). Local heating of the niobium film is performed with a tightly focused cw laser from which 40% of the optical power is absorbed. Vortex manipulation is performed by moving the laser beam with galvanometric mrrors (GM) placed in a telecentric system (TS). **b**, Magneto-optical image of a vortex lattice spontaneously created by field cooling of the niobium film down to $T$=4.6 K under $H_{ext}$= 0.024 Oe. **c**, Artificial vortex pattern engineered by single vortex repositionning from the initial vortex distribution displayed in b. The reposionning procedure is fully automatized, as described in the supplementary information. The laser with wavelength 532 nm is focused on the SC with a FWHM-diameter of 1.1 µm. The absorbed power is set to 17 µW. **d**, A new spontaneous vortex lattice is generated after a thermal cycle above $T_c$ and field cooling back to $T$=4.6 K under $H_{ext}$=1.64 Oe. **e**, From the intial vortex distribution displayed in d, a vortex-free area is produced by scanning the laser spot like a vortex-broom. The ending point of the spot trajectory is marked with a black cross. All scale bars are 20 µm.

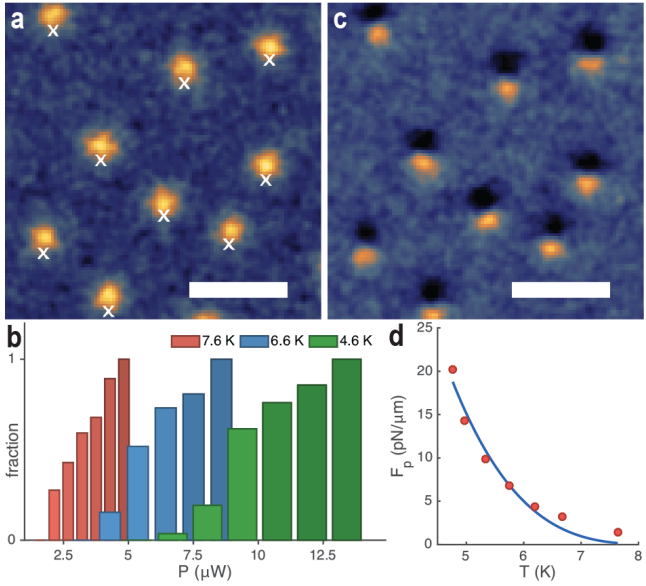

**Figure 1 : Laser power and temperature effects on vortex manipulation.**
**a**, Initial magneto-optical image of an area of the niobium film cooled at $T=4.6$ K in a magnetic field $\boldsymbol{H_{ext} = 0.22}$ Oe. A laser spot with a FWHM diameter of 1.1 μm is then successively placed at a fixed distance of 1.1 μm from each vortex. The central positions of the laser spot are marked with white crosses. **b**, Histograms of the fraction of untrapped vortices as a function of the absorbed laser power, for three different base temperatures of the SC. The statistics are built from 30 vortices. **c**, Image of the same area, built from the difference between MO-imaging contrasts after and before laser heating with an absorbed power of 13 μW. Thus only vortices that are untrapped and repositioned by the laser beam are visualized. In these conditions, all nine vortices have moved. **d**, Temperature dependence of the pinning force of a strongly bound vortex. The solid curve is a fit with the empirical power law $\boldsymbol{F_p \propto (1 - T/T_c)^\gamma}$, yielding the exponent $\boldsymbol{\gamma = 3.4}$. All scale bars are 10 μm.

In order to perform an efficient single vortex manipulation, it is crucial to chose a laser power low enough to keep the local temperature below $T_c$ and high enough so that the thermal force overcomes the pinning potentials. To determine the absorbed power needed to untrap all single vortices in a selected region of the SC at temperature $T$, we sequentially position the laser spot at a micrometric distance (~1μm) from each vortex (crosses in Fig. 2a) and count the number of untrapped vortices for various laser powers. As displayed in Fig 2b, the fraction of untrapped vortices strongly depends on the laser power and the base temperature, reflecting a large distribution of pinning potentials in the sample. Most notably, we demonstrate in Fig. 2c that 100 % of the vortices in the selected region can be dragged out of their pinning sites. Moreover, we checked that the superconductivity is not destroyed at the laser powers used for the manipulation (See Methods). This approach allows a straightforward determination of the pinning force of each vortex at any temperature. From Fig. 2b, we can estimate the pinning force of the most strongly bounded vortices in the selected SC region at various temperatures. At $T= 4.6$ K for instance, the thermal force overcomes all pinning forces for an absorbed power of 13 μW, corresponding to a temperature gradient at the vortex position of ~1.9 K/μm (see Methods). Using expression (2) of the thermal force and taking $\lambda_0 = 90$ nm and $\lambda_0/\xi_0 = 9$ relevant to our niobium film, we deduce a pinning force per vortex unit length $F_p\sim$19 pN/μm. This value is in accordance with previous determinations based on critical current measurements on similar samples[21-24]. Finally, the temperature dependence of the pinning force of a strongly bound vortex displayed is displayed in Fig 2d. It is well reproduced with the empirical power law $F_p \propto (1 - T/T_c)^\gamma$ with $\gamma = 3.4$, in agreement with previous ensemble measurements on vortices in niobium films[21,25].

The extreme simplicity of optical generation of strong local thermal gradients in SCs enables to tailor normal metal regions in the SC, and study the magnetic flux penetration close to their boundaries. For instance, starting from a spontaneous spatial distribution of vortices, a strong illumination of the Nb film with a focused laser produces a central region of radius $R_0$ where the temperature exceeds $T_c$. This normal (N) region thus traps a magnetic flux $\Phi \sim \pi H_{ext} R_0^2$ originating from the destroyed vortices (Fig. 3a). The averaged radial profile of the magnetic field during laser illumination is presented in Fig. 3c. It recalls that of a supercurrent loop of radius $R_0$ surrounding the N region (dashed blue curve in Fig 3c), which adds to the average magnetic field in the superconductor (defined as the total flux through the SC divided by the SC surface). After switching the laser off, the local temperature starts relaxing to its base value, inducing shrinkage of the N region where $T$ remains larger than $T_c$. The magnetic flux remains trapped within the N region as a result of geometrical barrier[26,27] that prevents the vortices from entering the superconducting region of the sample. To support this picture, we solve the heat equation to calculate the time evolution of the sample temperature profile at early times after the laser switch-off (red curves of Fig. 3d). For each profile, the solution of the stationary Ginzburg-Landau equation gives the corresponding spatial distribution of the normalized order parameter (blue curves) and therefore the size of the shrinking N region. Because of flux conservation, the trapped magnetic field increases with time (green levels) until it reaches the geometrical barrier critical field $H_p$ (penetration field) for a radius $R^* = R_0 \sqrt{H_{ext}/H_p}$. When the radius of the N region becomes smaller than $R^*$, the geometrical barrier vanishes and the vortices penetrate the SC region where they get trapped at the nearest pinning sites. This behavior explains the final distribution of flux quanta in figure Fig. 3b, characterized by a dense vortex region with radius $R^*$ (where the individual vortices are not optically resolved) belt by a vortex free SC region with an external radius $R_0$. Following the Bean critical state model, the vortex distribution is determined by the balance between the pinning force and the Lorentz driving force $j_c \Phi_0$, $j_c$ being a constant critical current density. This explains why the radial profile of the magnetic field presented in Fig 3e is almost linear in the vortex cluster. Interestingly, the experimental dependence of the ratio $(R^*/R_0)^2$ on the external magnetic field $H_{ext}$ plotted in Fig. 3f can be used to estimate the BL critical field (see Methods).

In conclusion, the performed experiments unveil new aspects of the interaction between laser radiation and the vortex matter. By choosing the appropriate laser parameters, one can realize various regimes of vortex manipulation, from the precise and rapid positioning of individual vortices to the generation of tight vortex bunches. The interplay between photons and single flux quanta should open up novel research directions in quantum computation based on braiding and entanglement of vortices, Josephson switches of electric current[28], or optically controlled elements of Rapid Single Flux Quantum logics[29] - a new research field that could be called "optofluxonics".

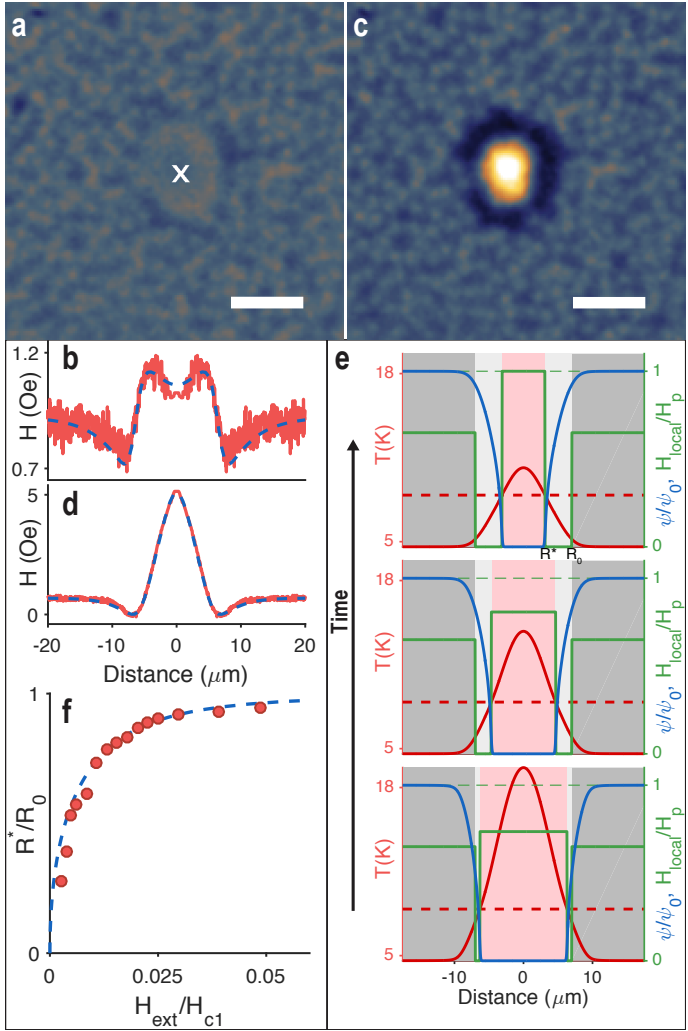

**Figure 3 : Dynamics of the magnetic flux penetration through a moving SC/N boundary.**

**a,** Magnetic field landscape during laser heating, for $H_{ext}$ = 3.0 Oe and a base SC temperature $T$=4.6 K. The central position of the laser spot is marked with a white cross. The absorbed power 450 μW sets a central region with radius $R_0$ = 6.5 μm in a N state. **b,** Averaged radial profile of the magnetic field. The theoretical profile (blue dashed curve) is obtained from the magnetic field created by a supercurrent loop surrounding the N region. **c,** After the laser switch-off, a dense vortex cluster with radius $R^*$ = 5.5 μm is surrounded by a vortex-free area with external radius $R_0$. **d,** Averaged radial profile of the magnetic field. The theoretical profile (blue dashed curve) is obtained from the Bean critical state model. **e,** Model of temporal evolution of the magnetic field penetration through the SC region after the laser swith-off. While the temperature profile (red curve) collapses, the profile of the order parameter (blue curve) tightens (see the Supplementary Information file). The magnetic field trapped in the N region increases (green levels) and will penetrate the SC region only when exceeding the geometrical barrier, i.e. for a N-region radius smaller than $R^*$. **f,** Dependence of the external magnetic field $H_{ext}/H_{c1}$ on the ratio $R^*/R_0$, where $H_{c1}$ is the critical field at zero temperature. The blue curve is deduced from a theoretical model of geometrical barrier developed in the Supplementary Information file. All scale bars are 10 μm.

# 1 Acknowledgements

We thank V. A. Skidanov and V. S. Stolyarov for providing the MO indicator and the niobium sample, respectively. We acknowledge the support by the European NanoSC COST Action MP1201, grants from the French National Agency for Research (ANR, project Electrovortex), Région Aquitaine, the French Ministry of Education and Research, the European Research Council and the Institut universitaire de France.

# 2 Author contributions

# 3 Additional information

## Methods

**Magneto-Optical (MO) imaging of vortices in a niobium film**

Magneto-optical imaging of individual vortices is based on the Faraday rotation of light polarization in a (MO) indicator placed onto the superconductor, in a crossed-polarizer beam path configuration. The superconductor is a niobium film of thickness 90 nm grown by magnetron sputtering on a silicon substrate (thickness 500 μm). The indicator is a 2.5 μm-thick Bi:LuIG garnet of composition $Lu_{3-x}Bi_xFe_{5-z}Ga_zO_{12}$ with x~0.9, z~1.0, with in-plane uniform saturation magnetization 50 G, which was grown by liquid phase epitaxy on a (100) oriented paramagnetic gadolinium-gallium-garnet $Gd_3Ga_5O_{12}$ substrate. This garnet has a high Verdet constant of ~ $0.06°μm^{-1}.mT^{-1}$ at light wavelengths around 530 nm.

The sample was mounted on a piezo scanner and inserted together with an aspheric lens (numerical aperture 0.49) in a closed-cycle helium cryostat. It was then submitted to a perpendicular external magnetic field $H_{ext}$ and cooled below the critical temperature $T_c = 8.6$ K in order to set the niobium film in the superconducting (SC) mixed state.

The MO contrast strongly depends on the extinction ratio of the ensemble polarizer-lens-crossed analyzer[1,2]. An extinction ratio ~$10^{-3}$ at cryogenic temperatures could be achieved. In order to enhance the contrast of MO imaging, a background subtraction procedure was used to suppress the contribution of defects at the sample surface and non-uniformity of the sample illumination. An image taken at $T > T_c$ under external magnetic field was subtracted from all raw images recorded at $T < T_c$ under the same applied magnetic field.

In extended data Figure 1, we show MO images of two vortices at different SC temperatures ranging from 4.1 K to 8.4 K. The optical resolution of vortex images at the lowest temperatures is ~ 2.5 μm, i.e. much larger than the diffraction limit (~0.6 μm). Indeed, it is set by the divergence of the magnetic field lines arising from the vortex since the Faraday rotation of light polarization is integrated in the whole indicator thickness. We see a slight increase in the vortex apparent size and a decrease in the MO contrast when the temperature is raised towards $T_c$.

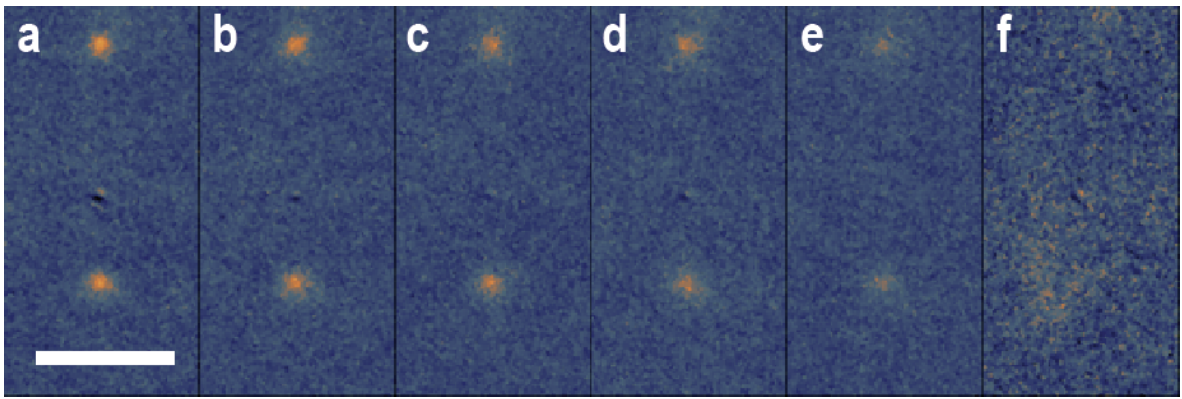

**Extended data Figure 1 : Vortex images at different SC temperatures.**
The vortices were created under an external magnetic field $H_{ext}$=0.03 Oe. **a**, T=4.1 K. **b**, T=6.6 K. **c**, T=7.1 K. **d**, T=7.6 K. **e**, T=8.1 K. **f**, T=8.4 K. The scale bar is 20 μm.

**Creating artificial vortex patterns**

Software was developed to locate all vortices of a selected area in a MO image and to relocate them to the new desired positions. An important point is to make sure that the laser beam manipulating a chosen vortex does not displace any other one during the whole vortex trajectory. The positions of the vortex centers are first identified with centroid calculations and used as seeds of a Voronoi diagram. The trajectory to reposition a single vortex from its initial location to its final position is then generated along segments of the Voronoi diagram so that the distance between the dragged vortex and the fixed vortices is maximized along the path.

**Estimation of the temperature profile in the SC**

The temperature profile created by a laser beam focused on our SC sample is calculated numerically using COMSOL program.
This program solves the stationary heat equation taking into account temperature dependent thermal conductivities of the SC and the substrate (for instance 0.05 W.m$^{-1}$.K$^{-1}$ at T=4.6 K for niobium[3] and 450 W.m$^{-1}$.K$^{-1}$ at T=4.6 K for Si). We consider a Gaussian shaped heat source applied to the top of the niobium film:

$$Q(r) = \frac{P}{2\pi r_0^2} \exp\left[\frac{-r^2}{2r_0^2}\right]$$

where $P$ is the total power absorbed by the SC and $r_0$ the size of the laser spot.

Extended data Figure 2 shows the profiles of the temperature and the temperature gradient calculated for $P$= 13 µW, $r_0$=0.5 µm and a base temperature $T$ = 4.6 K, matching the experimental conditions of Fig. 2. Clearly the whole temperature profile is below $T_c$.

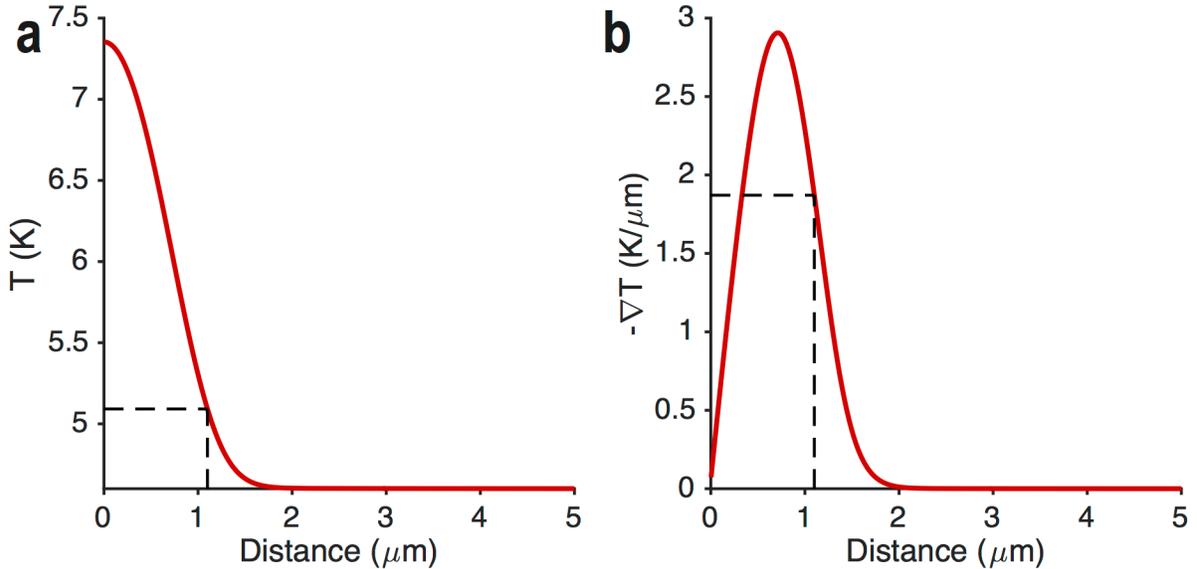

**Extended data Figure 2 : Temperature profile in the niobium film under laser heating.**
For a vortex located at a distance r=1.1 µm from the center of the laser spot, the thermal gradient is estimated to 1.9 K.µm$^{-1}$, which leads to a thermal force per vortex unit length F = 19 pN/µm according to Eq. 2, taking $\xi_0$ = 10 nm and $\lambda_0$ = 90 nm.

Moreover, we can experimentally demonstrate that single vortices can be safely manipulated with a focused laser that does not destroy superconductivity. Extended data Fig. 3a, b, c display the MO images of a strongly pinned vortex (named A) before, during and after illumination with a laser beam focused 1 μm away from the initial vortex position. The absorbed power is set to 17 μW, high enough to untrap this vortex and drop it at a new pinning site. Clearly, the image of vortex A is almost unaffected under illumination, meaning that the local temperature remains well below $T_c$ when performing single vortex optical manipulation.

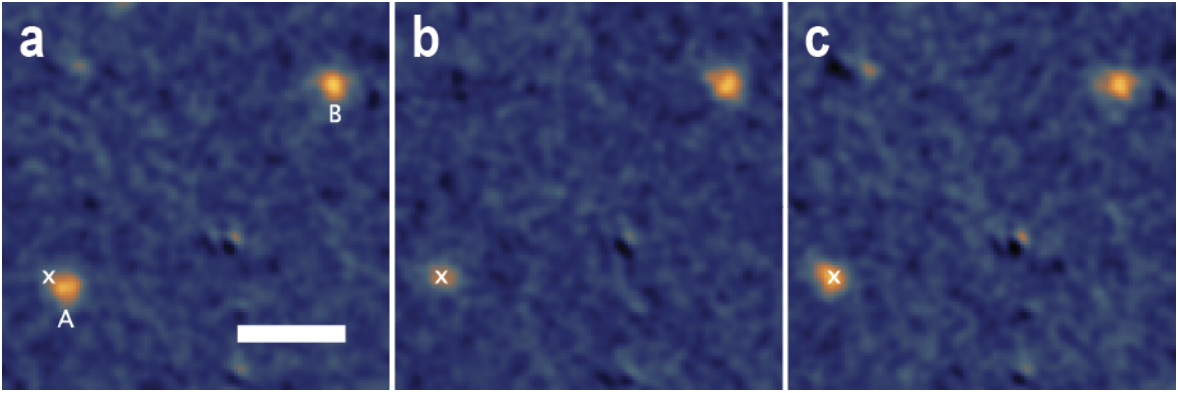

**Extended data Figure 3 : vortex survival under laser illumination.**
**a**, Image of a vortex strongly bound to a pinning site (vortex A), at $T$=4.6 K. Vortex B is a reference vortex. **b**, Image of the same area under laser heating at the location marked with a white cross, with an absorbed power of 17 μW. **c**, Image of the same area after laser heating. The scale bar is 10 μm.

**Calculation of the order parameter profile across the N/SC boundary**

To study the evolution of vortex distribution after heating the SC above $T_c$ with strong illumination we solve the Ginzburg-Landau (GL) equation to derive the spatial distribution of the order parameter for each temperature profile during the thermal relaxation.

Since the relaxation time of the order parameter $\psi$ is much smaller than that of the temperature profile, we consider an adiabatic evolution of $\psi$, i.e $\partial_t \psi = 0$. Moreover, considering that the scale of the order parameter variation is much smaller than the hot-spot characteristic size, the term with the first derivative $\partial_r \psi$ can be neglected in $\Delta \psi$, so that the Ginzburg-Landau equation writes

$$-\frac{\hbar^2}{4m}\frac{\partial^2\psi}{\partial r^2} + \alpha[T(r) - T_c]\psi + b\psi^3 = 0 \ .$$

Far away from the laser induced hot spot, the temperature of the SC is equal to $T_0 < T_c$ and the order parameter $\psi_\infty$ is uniform, determined by $\psi_\infty^2 = \alpha(T_c - T_0)/b$. Substituting the dimensionless order parameter $\Psi = \psi/\psi_\infty$ into the Ginzburg-Landau equation, we get:

$$-\frac{\hbar^2}{4m}\frac{\partial^2\Psi}{\partial r^2} + \alpha[T(r) - T_c]\Psi + \alpha(T_c - T_0)\Psi^3 = 0$$

After division by $\alpha(T_c - T_0)$ and introduction of the superconducting coherence length $\xi_0$ defined by $\xi_0^2 = \frac{\hbar^2}{4m\alpha T_c}$, the equation writes

$$-\frac{\xi_0^2}{1 - T_0/T_c}\frac{\partial^2\Psi}{\partial r^2} + \frac{T(r) - T_c}{T_c - T_0}\Psi + \Psi^3 = 0,$$

with the boundary conditions $\psi(r=0)=0$ and $\psi(r\rightarrow\infty)=1$.

The dimensionless form of this equation

$$\frac{\partial^2\psi}{\partial\rho^2}=\tau(\rho)\Psi+\Psi^3, \text{ with } \rho=\frac{r}{\xi_0}\sqrt{1-\frac{T_0}{T_c}} \text{ and } \tau(\rho)=\frac{T(\rho)-T_c}{T_c-T_0},$$

was numerically solved. The profile of the order parameter just after the laser switch-off is displayed in Extended data Figure 4a (blue curve).

Analytical solutions can be derived close to the circular N/SC moving boundary which is at a radius $R$. In this region the temperature is close to $T_c$ and can be expressed using the expansion:

$$T(r)\approx T_c\left[1-\frac{|\nabla T|_{r=R}}{T_c}(r-R)\right]$$

Defining the thermal length $L_T=T_c/|\nabla T|_{r=R}$, the GL equation takes the form:

$$-\xi_0^2\frac{\partial^2\Psi}{\partial r^2}-\frac{(r-R)}{L_T}\Psi+\left(1-\frac{T_0}{T_c}\right)\Psi^3=0$$

i) In the region where $r-R\lesssim\xi_0$, the non-linear term can be neglected ($\Psi^3\ll\Psi$) and the equation can be approximated by

$$\frac{\partial^2\Psi}{\partial r^2}+\frac{r-R}{L_T\xi_0^2}\Psi=0$$

The order parameter is therefore proportional to the Airy function $Ai\left[-\frac{r-R}{L_\psi}\right]$, where $L_\Psi=(\xi_0^2L_T)^{\frac{1}{3}}$ is a characteristic thickness of the order parameter front.

ii) In the region where $r-R>\xi_0$, the typical scale of the order parameter variation is much larger than $\xi_0$ and the term $\xi_0^2\partial_r^2\Psi$ in the GL equation can be neglected. The order parameter thus evolves as $\sqrt{(r-R)/L_T}$. As shown in Extended data Fig. 4b, the numerical profile of the order parameter at the boundary region can be well approximated with the analytical functions derived above.

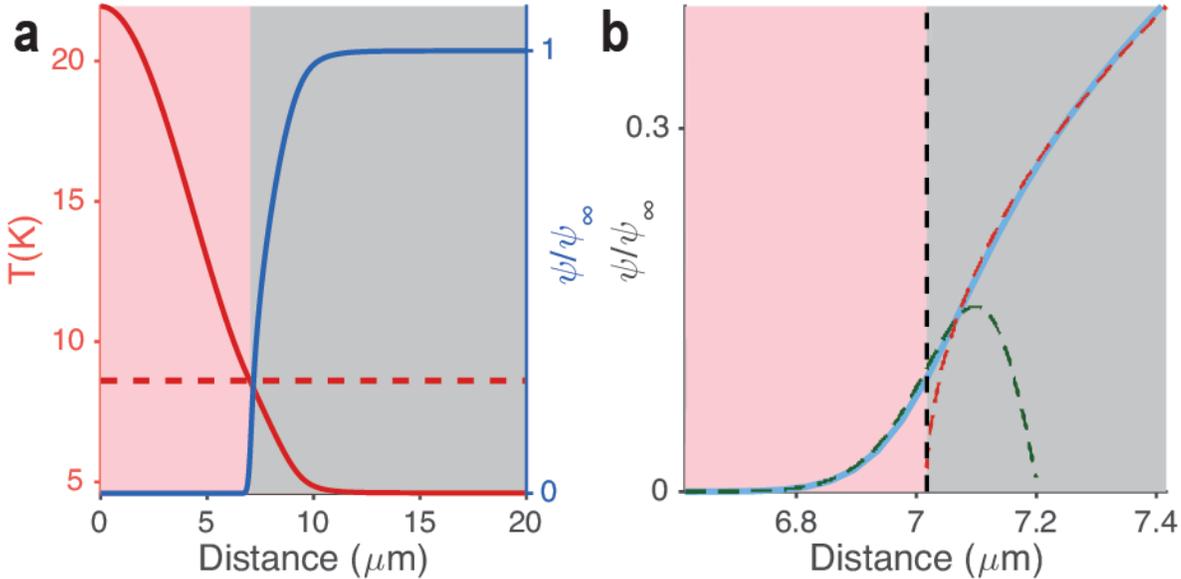

**Extended data Figure 4: Calculated order parameter profile across N/SC boundary**
**a**, Red curve : calculated temperature profile just after the laser switch-off. The critical temperature of the niobium film (dashed level) is reached for a radius $R_0 = 7$ µm. Blue curve : numerical solution of the GL equation.
**b**, Blue curve : Zoom of the numerical solution of the GL equation at the N/SC boundary. Dashed curves : Analytical solutions of the Ginzburg-Landau equation for $R = R_0$ at the foot of the order parameter front (green, Airy function comprising an exponential growth of $\Psi$) and in the SC region (red, square root evolution of $\Psi$).

**Estimation of the geometrical barrier**

The presence of the order parameter front provides a natural explanation for the vortex-free region surrounding the dense vortex cluster. After switching off the laser, the temperature starts to decrease and the N region shrinks. However, the magnetic flux stays trapped inside the N region until the magnetic field reaches the geometrical barrier critical value. The main physical idea of the geometrical barrier, introduced in Ref.[7] may be also applied to our case. Since the superconducting film thickness $d$ is smaller than the London penetration depth $\lambda$, the magnetic field spreads from the vortex core over a distance $\lambda_{eff} = \lambda^2/d$ larger than $\lambda$ (see Ref.[8]). In this situation, the characteristic distance $\zeta$ from the N/S interface at which the main variations of the vortex energy will be $\lambda_{eff}$. In our case, $\lambda$ strongly varies with the distance from the N /S boundary because of the temperature profile. Thus, the characteristic distance $\zeta$ should be estimated self-consistently by writing $\lambda_{eff}(T) \sim \zeta$. Taking $\lambda \cong \lambda_0 \sqrt{L_T/\zeta}$, we obtain $\zeta \sim \lambda_0 \sqrt{(L_T/d)}$.

As described in the Ref.[7], vortex penetration occurs when the Lorentz force $J\phi_0$ induced by the supercurrent density $J$ reaches the value of $2\varepsilon_0/\zeta$, where $2\varepsilon_0 = \frac{\phi_0}{2\pi} \frac{H_{c1}(0)\zeta}{L_T}$ is the vortex line energy per unit length. This leads to :

$$J\phi_0 = \frac{2\phi_0 H_{c1}(T=0)}{4\pi L_T}$$

By adapting the expression of the Meissner current $J$ of Ref.[4] to our ring-shaped strip of width $R_0 - R^*$, which is valid in the limit $|R_0 - R^*| \ll R_0$, we obtain close to the N/S boundary and when the local field reaches the penetration field $H_p$:

$$J(R^* + \zeta) \cong \frac{H_p}{4\pi d} \sqrt{\frac{R_0 - R^*}{\zeta}}$$

This leads to the following expression for $H_p$:

$$H_p = 2 H_{c1}(T=0) \sqrt{\frac{\zeta}{(R_0 - R^*)}} \frac{d}{L_T} = 2 H_{c1}(T=0) \sqrt{\frac{d}{(R_0 - R^*)}} \left(\frac{\lambda_0^2 d}{L_T^3}\right)^{\frac{1}{4}}$$

The resolution of the heat equation leads to the temperature profile in our experimental conditions and yields $L_T \sim 5.3$ µm. Taking $R_0 \sim 6.5$ µm and $R^* \sim 5.5$ µm (Fig. 3b-d), we obtain a penetration field $H_p \sim 6$ $Oe$, which is in good agreement with the estimation of the local field $(R_0/R^*)^2 H_{ext} = 4.3$ $Oe$ from the experimental data. The relation between the external magnetic field and the geometrical parameters of the vortex-free region is given by:

$$H_{ext} = 2H_{c1}(0) \left(\frac{R^*}{R_0}\right)^2 \sqrt{\frac{d}{(R_0 - R^*)}} \left(\frac{\lambda_0^2 d}{L_T^3}\right)^{\frac{1}{4}}$$

To adjust the experimental data of figure dependence of Fig. 3f, we thus used the formula

$$\frac{H_{ext}}{H_{c1}(0)} = \frac{x^2}{\sqrt{1-x}} \left(\frac{16\lambda_0^2 d^3}{R_0^2 L_T^3}\right)^{\frac{1}{4}} \cong 0.01 \frac{x^2}{\sqrt{1-x}}$$

## Modelization of the magnetic field profile in the vortex cluster

Following the Bean critical state model, the vortex distribution is determined by the balance between the pinning force and the Lorentz driving force $j_c \Phi_0$, which pushes the vortices toward the SC region[9]. The radial profile of the vertical component of magnetic field, built from $j_c = -\frac{\partial H_z}{\partial r}$, is therefore linear in a thick superconductor. In order to match our experimental conditions, we consider here a SC film with finite thickness $d$, which occupies the region $-d < z < 0$. The current distribution in the vortex cluster with radius $R^*$ can be modeled as

$$j_\theta = \begin{cases} j_c & \text{for} \quad r < R^* \\ 0 & \text{for} \quad r > R^* \end{cases}.$$

The vertical component of magnetic field created by this distribution in the plane $z = 0$ and at a radial distance $r$ is given by:

$$H_z(r,0) = \frac{j_c}{2\pi} \int_{-d}^{0} dz_0 \int_{0}^{R^*} \frac{1}{\sqrt{(a+r)^2 + z_0^2}} \left[ K\left(\frac{4ar}{(a+r)^2 + z_0^2}\right) + \frac{a^2 - r^2 - z_0^2}{(a-r)^2 + z_0^2} E\left(\frac{4ar}{(a+r)^2 + z_0^2}\right) \right] da ,$$

where K and E are full elliptic integrals of the first and second kind, respectively[10].

In order to match the experimental observations, we add to $H_z$ a contribution which takes into account the averaged magnetic field $H_v$ arising from the vortices imaged at $r > R_0$. This field is modeled with

$$H_v = \frac{H_{av}}{2} \left( 1 + erf\left(\frac{r - R_0}{\delta}\right) \right)$$

where $H_{av}$ is a field amplitude and $\delta$ is the width of the magnetic field step, corresponding to our MO resolution. The experimental magnetic field profile is well reproduced by summing $H_z$ and $H_v$ (see Fig. 3d), taking $\delta = 3.3$ μm, $R_0$=7.5 μm, $H_{av}$=0.9 Oe.